\documentclass[conference,letterpaper]{IEEEtran}

\usepackage{cite}
\usepackage{fancyhdr}
\usepackage{graphicx}
\usepackage{tikz}
\usepackage{amsmath}
\usepackage{amssymb}
\usepackage{booktabs}  
\usepackage{paralist}
\newcommand{\para}[1]{\smallskip\noindent\textbf{{#1.}}}
\usepackage{minted}
\usepackage[utf8]{inputenc}    
\usepackage{enumitem}    
\usepackage{comment}
\usepackage{subcaption}
\usepackage{url}           
\usepackage{hyperref}      
\usepackage{cleveref}
\usepackage{tabularx}      
\usepackage{longtable}     
\usepackage{makecell}      
\usepackage{textcomp}      
\usepackage{ragged2e}      
\usepackage{lipsum}
\usepackage{threeparttable}
\usepackage[table]{xcolor}
\crefname{algocf}{algorithm}{algorithms}
\Crefname{algocf}{Algorithm}{Algorithms}
\usepackage[T1]{fontenc} 
\usepackage{iftex} 
\usepackage{caption} 
\usepackage[inkscapelatex=false]{svg} 

\newcolumntype{L}{>{\RaggedRight\arraybackslash}X}
\newcolumntype{C}{>{\centering\arraybackslash}X}

\definecolor{tablehead}{HTML}{442266}

\renewcommand{\figurename}{Fig.}

\newcommand{\param}[1]{\textcolor{red}{#1}} 
\newcommand{\xxx}[1]{\param{XXX}} 

\newcommand{\ignore}[1]{}

\newcounter{obs}
\definecolor{nbs}{rgb}{0.88, 0.07, 0.37}
\definecolor{agyc}{rgb}{0.37, 0.88, 0.07}
\definecolor{moegi}{rgb}{0.357, 0.537, 0.188}
\definecolor{burntorange}{rgb}{0.8, 0.33, 0.0}
\definecolor{carmine}{rgb}{0.59, 0.0, 0.09}
\definecolor{ceruleanblue}{rgb}{0.16, 0.32, 0.75}

\newif\ifcamerareadyiterations
\camerareadyiterationsfalse

\newif\ifcameraready
\camerareadytrue

\newif\ifdraft
\draftfalse

\newif\ifblind
\blindtrue

\newif\ifispassdraft
\ispassdraftfalse

\newif\ifshepherd
\shepherdfalse

\ifshepherd
\usepackage[colorinlistoftodos,prependcaption,textsize=small]{todonotes}

\newcommand{\scomment}[2]{\todo[size=\scriptsize, linecolor=ceruleanblue, bordercolor=ceruleanblue, backgroundcolor=white!90!ceruleanblue]{\textcolor{ceruleanblue}{
\textbf{S#1} #2}}}
\newcommand{\rcomment}[1]{\todo[size=\scriptsize, linecolor=burntorange, bordercolor=burntorange, backgroundcolor=white!90!burntorange]{\textcolor{burntorange}{
\textbf{#1}}}}

\else

\newcommand{\scomment}[2]{}
\newcommand{\rcomment}[1]{}

\fi

\ifdraft
    \usepackage[colorinlistoftodos,prependcaption,textsize=small]{todonotes}
    \newcommand{\nbcomment}[1]{\todo[size=\scriptsize, linecolor=orange, bordercolor=orange, backgroundcolor=white]{\textcolor{nbs}{\textbf{@nb:} #1}}}

    \newcommand{\mma}[1]{\textcolor{orange}{\textbf{[@mma: #1]}}}
    \newcommand\mmatodo[1]{{\textcolor{orange}{#1}}}
    \newcommand{\atbcomment}[1]{\todo[size=\scriptsize, linecolor=orange, bordercolor=orange, backgroundcolor=white]{\textcolor{blue}{\textbf{@atb:} #1}}}

    \newcommand{\omcrcomment}[1]{\todo[size=\scriptsize, linecolor=orange, bordercolor=orange, backgroundcolor=white]{\textcolor{blue}{\textbf{@om:} #1}}}

\else 
    \newcommand{\nbcomment}[1]{}

    \newcommand{\mma}[1]{}
    \newcommand\mmatodo[1]{}
    \newcommand\atbcomment[1]{}
    \newcommand\mmacomment[1]{}

    \newcommand{\omcrcomment}[1]{}
\fi

\ifispassdraft
    \usepackage[colorinlistoftodos,prependcaption,textsize=small]{todonotes}
    \renewcommand{\nbcomment}[1]{\todo[size=\scriptsize, linecolor=orange, bordercolor=orange, backgroundcolor=white]{\textcolor{nbs}{\textbf{@nb:} #1}}}
    
\fi

\definecolor{aliceblue}{rgb}{0.94, 0.97, 1.0}

\usepackage[framemethod=tikz]{mdframed}

\usepackage[most]{tcolorbox} 
\newtcolorbox[auto counter]{tkx}[2][]{%
    enhanced, breakable, center title,
    colframe = #2!45,
    colback  = #2!10,
    colbacktitle=#2!20,
    left=1.5pt,
    right=1.5pt,
    bottom=1.5pt,
    top=1.5pt,
    #1%
}

\newcounter{tkw}
\newcommand\takeaway[1]{
\stepcounter{tkw}
\begin{tkx}{Maroon}
\noindent\small{\textbf{SPEC's Warning \cite{warning}:} #1}
\end{tkx}
}

\PassOptionsToPackage{hyphens}{url}\usepackage{hyperref}
\usepackage[dvipsnames,svgnames]{xcolor} 
\hypersetup{
    colorlinks=true,
    linkcolor=ForestGreen,
    citecolor=ForestGreen,
    urlcolor=RoyalBlue,
    filecolor=IndianRed,
    pdftitle={Adaptation Fidelity of SPEC CPU 2026}, 
    pdfauthor={Doa'a Al-Otoom, Mahesh Madhav}, 
    pdfsubject={Benchmarks}, 
    pdfkeywords={Benchmarks, CPU Performance}, 
    pdfnewwindow=true,
    pdfdisplaydoctitle=true,
    bookmarksopen=false
}

\begin{document}

\makeatletter
\newcommand\HUGE{\@setfontsize\Huge{27}{27}}
\makeatother  

\title{\HUGE{Adaptation Fidelity of SPEC CPU$^\circledR$2026}}

\author{}
\author{%
  \IEEEauthorblockN{Doa'a Al-Otoom, Mahesh Madhav}
  \IEEEauthorblockA{\emph{Ampere Computing, Portland, OR}}
}

\maketitle

\newcommand{\Red}[1]{{\color{red} #1}}
\newcommand{\asm}[1]{\texttt{#1}}
\newcommand{\sys}[1]{\texttt{#1}}
\newcommand{\kw}[1]{\textit{#1}}
\newcommand{\kwb}[1]{\textbf{#1}}
\newcommand{\type}[1]{\textit{#1}}
\newcommand{\Response}[1]{{\color{blue} #1}}
\newcommand{\XXX}[1]{\Red{\textbf{XXX[}#1\textbf{]}}}
\newcommand{\tocite}[1]{\Red{CITE:\cite{#1}}}
\newcommand{\toref}[1]{\Red{REF:\ref{#1}}}
\newcommand{\parasub}[1]{\smallskip\noindent\textit{{#1:}\xspace}}
\newcommand{\mahesh}[1]{\textcolor{purple}{#1}}
\newcommand{\anyone}[1]{\textcolor{blue}{#1}}
\newcommand{\niparagraph}[1]{\noindent\textbf{\textsf{#1}\hspace{0.5em}}}
\newcommand\TODO[1]{\textcolor{red}{TODO: #1}}

\newenvironment{CompactItemize}%
  {\begin{list}{$\blacktriangleright$}%
    {\leftmargin=\parindent \itemsep=2pt \topsep=2pt
     \parsep=0pt \partopsep=0pt}}%
  {\end{list}}
\renewcommand{\labelitemi}{$\blacktriangleright$}

\newcommand{\malloc}{{\texttt{Malloc}}}
\newcommand{\linklist}{{\textsf{Linked-List}}}
\newcommand{\plusplus}{\texttt{++}}

\begin{abstract}
Standardized benchmarks are often criticized for not being ``real workloads,'' but this critique is rarely backed by data. This paper provides the first systematic, quantitative analysis of the ``fidelity gap'' between the SPEC CPU\textregistered{}2026 suite and its original, upstream open-source counterparts. We compile both the SPEC benchmarks and their upstream applications and execute them with official input workloads under two scenarios: a single-copy latency run and a 192-copy throughput run.

Our findings show that most benchmarks exhibit high fidelity in single-copy runs, while a few outliers reveal the impact of SPEC's adaptation process. The multi-copy results further highlight the necessity of this adaptation: several benchmarks become significantly more efficient than their upstream versions under heavy load, underscoring the importance of I/O reduction. This work offers data-driven validation of SPEC's methodology, showing that the fidelity gap is not a flaw but a quantifiable consequence of enforcing portability, determinism, and CPU-centric measurement.
\end{abstract}

\section{Introduction}

Standardized benchmarks are the bedrock of quantitative research in computer architecture, yet their very nature creates a philosophical dilemma. To be effective, they must be derived from ``real workloads,'' a principle the SPEC CPU committee embraces by founding each benchmark on a real-world application and preserving its essential performance characteristics \cite{how_to_spec}. To be scientific, however, they must be transformed into deterministic, portable, and reproducible instruments \cite{cpu2026}. This transformation, which removes randomness, I/O, and platform-specific code, is essential for fair CPU comparisons but raises an important question: how much realism is sacrificed for the sake of rigor?

This question is often at the heart of critiques that dismiss benchmarks as ``synthetic'' \cite{amazon_reinvent, reddit, statsig}. While these concerns are valid, they are rarely substantiated with quantitative evidence; the fidelity gap between an application and its benchmark derivative has remained a subject of speculation rather than systematic study. While there have been numerous studies on how representative SPEC CPU is of industry workloads \cite{spec_499, cpu2017_broaden} or other benchmarks \cite{cpu2026_comparison}, none have focused on the loss of fidelity from the adaptation process itself.

This paper provides the first comprehensive, empirical investigation into this fidelity gap for SPEC CPU\textregistered{}2026. For each of the 26 benchmarks in SPECrate, we compile its original upstream open-source counterpart and measure its performance against the official benchmark version using identical input workloads, precisely quantifying the delta introduced by adaptation. By analyzing these deltas, we identify which benchmarks serve as high-fidelity proxies for their in-field counterparts and which have been more significantly perturbed; for the latter, we attribute the divergence to specific modifications such as the removal of intrinsics or I/O operations.

Ultimately, this paper offers a more nuanced, data-driven understanding of where SPEC CPU 2026 accurately reflects real-world performance and where its role as a standardized instrument causes it to diverge. The following sections detail our methodology, present our findings, and discuss the implications for benchmark users and developers alike.

\para{Sentinel at the Gate}
The SPEC CPU documentation clearly states that benchmarks adapted from applications may not look and feel like the original application: 

\takeaway{The individual benchmarks in this suite may be similar, but are NOT identical to benchmarks or programs with similar names available from sources other than SPEC. SPEC has invested significant effort to improve portability and minimize hardware dependencies, ... so the application programs in this distribution may perform differently from commercial or open-source versions of the same application. Therefore, it is not valid to compare SPEC CPU 2026 results with anything other than other SPEC CPU 2026 results.}

\noindent This adaptation process is even laid out in detail in the SPEC CPU 2026 announcement paper \cite{cpu2026}. Nevertheless, we throw caution to the wind and roll up our sleeves...

\section{Methodology}

To empirically quantify the fidelity gap between the originally snapshotted open-source applications and their adapted benchmark counterparts, we designed a controlled experiment centered on building two sets of binaries from a common source baseline, executing them in an identical environment, and verifying the results.

\subsection{Build and Execution}
First, we compile the ``Upstream Application'' binaries from the ``redistributable source'' package provided by SPEC for each benchmark, which is the code snapshot that SPEC started with. Second, we compile the official SPEC binaries, the ``Adapted Benchmark,'' directly using the SPEC CPU harness after the full adaptation and code-hardening process (removal of I/O, non-determinism, and platform-specific code). Both are built with \texttt{gcc-15.2} and \texttt{-O3}, based on the example gcc configuration offered by SPEC.

To isolate code changes as the only variable, we modify each benchmark's \texttt{speccmds.cmd} file to replace the call to the adapted binary with a call to our compiled upstream binary, then execute it via \texttt{specinvoke} with the same input workloads, command-line arguments, and environment variables as the official run. A few exceptions required deviation from this process; these are noted in \autoref{notes}.

All experiments were conducted on AArch64. While SPEC benchmarks are designed for portability, many upstream counterparts are developed on x86, so porting to a non-primary architecture serves as a useful stress test and can amplify fidelity gaps related to system interactions, since system call overhead can be higher on AArch64 \cite{serverless}.

Performance was measured under two load scenarios, both plotted as S-curves sorted by speedup \autoref{fig:1copy_results}, \autoref{fig:192copy_results} and summarized in \autoref{tab:speedup-data}. The single-copy run uses the official refrate workload size (the SPECrate submission standard) and measures elapsed runtime for a single invocation of each binary, capturing performance under minimal system load. The multi-copy run scales this to a 192-copy SPECrate execution on a 192-core processor, amplifying system-level effects such as residual I/O or shared-resource contention that a single-copy run may hide.

\subsection{Verification}
A core tenet of the SPEC methodology is not just to measure how fast a workload runs, but how fast it runs correctly. To uphold this principle and ensure that the upstream applications performed an equivalent quantum of work, we integrated the official SPEC verification process into our experiment. After each run of the upstream application binary, we execute the \texttt{specinvoke compare.cmd} command provided by the harness. This command validates the application's output against the golden reference output files, checking for correctness within the specified tolerances for each benchmark. In the vast majority of cases the upstream output matched the reference, confirming equivalent work; discrepancies were analyzed manually (e.g., in 748.flightdm, SPEC removed redundant configuration dumps and verbose step logging; in 729.abc, SPEC removed a build timestamp printout). This step ensures measured deltas reflect execution efficiency rather than differences in work performed.

\subsection{Platform}
All measurements were conducted on an AArch64 platform based on the AmpereOne SoC \cite{AmpereOne_Brief}, configured as shown in \autoref{tab:sys-config}. The results shown in this paper are reproducible on this platform; any other platform will behave differently and may even report opposite results compared to what is observed below. We offer root causes for the performance discrepancies we observed, summarized in \autoref{tab:speedup-data} with outliers detailed in \autoref{notes}.
\begin{table}[ht]
\footnotesize
\centering
\setlength{\aboverulesep}{0pt}
\setlength{\belowrulesep}{0pt}
\setlength{\extrarowheight}{.75ex}

\caption{Experimental System Configuration}
\label{tab:sys-config}

\begin{tabularx}{\dimexpr\columnwidth - 8pt\relax}{@{}
  >{\hsize=0.7\hsize}L
  >{\hsize=1.30\hsize}L
  @{}}
\arrayrulecolor{lightgray!40}

\rowcolor{tablehead}
\thead[l]{\textbf{\textcolor{white}{Component}}} &
\thead[l]{\textbf{\textcolor{white}{Configuration}}} \\

\bottomrule
Processor &  AmpereOne A192-32X \cite{AmpereOne_Brief} \\
\bottomrule
Cores & 192 Ampere Cores (aarch64, ARMv8.6) \\
\bottomrule
Frequency & 3.2 GHz \\
\bottomrule
Memory &  1.0 TB DDR5-5600 \\
\bottomrule
L1 Cache &  16 KB code + 64 KB data \\
\bottomrule
L2 Cache &  2 MB \\
\bottomrule
L3 Cache &  64 MB SLC \\
\bottomrule
Compiler and Flags &  GCC 15.2 -O3 \\
\bottomrule
Operating Environment &  Ubuntu 24.04.4 LTS \\& Linux kernel: 6.8.0-110-generic-64k \\
\bottomrule
NUMA &  1 Node per socket (Monolithic)\\
\bottomrule
\end{tabularx}
\par
\setlength{\aboverulesep}{0.6ex} 
\setlength{\belowrulesep}{0.9ex} 
\setlength{\extrarowheight}{0pt}
\end{table}

\section{Results}
\label{sec:results}

\definecolor{low}{HTML}{a31111} 
\definecolor{mid}{HTML}{FFFFFF}
\definecolor{high}{HTML}{118311} 
\newcommand*{\opacity}{90}

\newcommand*{\minval}{0.500}
\newcommand*{\midval}{1.000}
\newcommand*{\maxval}{2.000}

\newcommand{\gradient}[1]{
    \ifdimcomp{#1pt}{>}{\maxval pt}{#1}{
        \ifdimcomp{#1pt}{<}{\minval pt}{#1}{
            \ifdimcomp{#1pt}{<}{\midval pt}{
                \pgfmathparse{int(round(100*(#1-\minval)/(\midval-\minval)))}
                \xdef\tempa{\pgfmathresult}
                \cellcolor{mid!\tempa!low!\opacity} #1
            }{
                \pgfmathparse{int(round(100*(#1-\midval)/(\maxval-\midval)))}
                \xdef\tempa{\pgfmathresult}
                \cellcolor{high!\tempa!mid!\opacity} #1
            }
            
    }}
}
\newcommand{\gradientbold}[1]{
    \ifdimcomp{#1pt}{>}{\maxval pt}{#1}{
        \ifdimcomp{#1pt}{<}{\minval pt}{#1}{
            \ifdimcomp{#1pt}{<}{\midval pt}{
                \pgfmathparse{int(round(100*(#1-\minval)/(\midval-\minval)))}
                \xdef\tempa{\pgfmathresult}
                \cellcolor{mid!\tempa!low!\opacity} \textbf{#1}
            }{
                \pgfmathparse{int(round(100*(#1-\midval)/(\maxval-\midval)))}
                \xdef\tempa{\pgfmathresult}
                \cellcolor{high!\tempa!mid!\opacity} \textbf{#1}
            }            
    }}
}

\begin{table}[!t]
\footnotesize
\centering
\setlength{\aboverulesep}{0pt}
\setlength{\belowrulesep}{0pt}
\setlength{\extrarowheight}{.75ex}

\caption{Performance comparison table. This is a summary of the data in Figures \ref{fig:1copy_results} and \ref{fig:192copy_results}. Green means SPEC is faster, red means SPEC is slower.}
\label{tab:speedup-data}

{\centering

\begin{tabularx}{\dimexpr\columnwidth - 8pt\relax}{@{}
  >{\hsize=1\hsize}L
  >{\hsize=.7\hsize}C
  >{\hsize=.7\hsize}C
  >{\hsize=1.6\hsize}L
  @{}}

\arrayrulecolor{lightgray!40}
\rowcolor{tablehead}
\thead[l]{\rule{0pt}{1.6ex}\textbf{\textcolor{white}{benchmark}}} &
\thead[c]{\rule{0pt}{1.6ex}\textbf{\textcolor{white}{1-copy}}} &
\thead[c]{\rule{0pt}{1.6ex}\textbf{\textcolor{white}{192-copy}}} &
\thead[l]{\rule{0pt}{1.6ex}\textbf{\textcolor{white}{causes of difference}}} \\
\midrule

\textbf{706.stockfish} & \gradient{0.615} & \gradient{0.755} & intrinsics removal \\ \hline
\textbf{707.ntest} & \gradient{1.032} & \gradient{1.711} & i/o efficiency \\ \hline
\textbf{708.sqlite} & \gradient{1.442} & \gradient{1.309} & changed fp types \\ \hline
\textbf{709.cactus} & \gradient{1.013} & \gradient{1.015} &  \\ \hline
\textbf{710.omnetpp} & \gradient{1.078} & \gradient{1.047} & threading overhead \\ \hline
\textbf{714.cpython} & \gradient{0.944} & \gradient{0.964} & no inlines \\ \hline
\textbf{721.gcc} & \gradient{1.140} & \gradient{1.031} & system time reduction \\ \hline
\textbf{722.palm} & \gradient{1.033} & \gradient{1.001} \\ \hline
\textbf{723.llvm} & \gradient{1.076} & \gradient{1.361} & i/o write reduction \\ \hline
\textbf{727.cppcheck} & \gradient{0.978} & \gradient{1.001} \\ \hline
\textbf{729.abc} & \gradient{0.985} & \gradient{0.962} \\ \hline
\textbf{731.astcenc} & \gradient{0.642} & \gradient{0.643} & intrinsincs removal \\ \hline
\textbf{734.vpr} & \gradient{0.974} & \gradient{0.921} & no inlines, stable sort \\ \hline
\textbf{735.gem5} & \gradient{1.113} & \gradient{1.378} & i/o write reduction \\ \hline
\textbf{736.ocio} & \gradient{1.001} & \gradient{0.998} \\ \hline
\textbf{737.gmsh} & \gradient{1.029} & \gradient{1.006} \\ \hline
\textbf{748.flightdm} & \gradient{1.012} & \gradient{1.017} & i/o efficiency \\  \hline
\textbf{749.fotonik3d} & [N/A] & [N/A] & no upstream code \\ \hline
\textbf{750.sealcrypto} & \gradient{1.042} & \gradient{1.034} \\ \hline
\textbf{753.ns3} & \gradient{1.018} & \gradient{1.037} & self-timing removal \\ \hline
\textbf{765.roms} & \gradient{0.974} & \gradient{1.000} \\ \hline
\textbf{766.femflow} & \gradient{1.176} & \gradient{1.077} & threading overhead \\ \hline
\textbf{767.nest} & \gradient{1.179} & \gradient{1.052} & RNG, thread overheads \\ \hline
\textbf{772.marian} & [N/A] & [N/A] & does not build on arm\\ \hline
\textbf{777.zstd} & \gradient{0.943} & \gradient{0.972} &  \_\_builtin removals \\ \hline
\textbf{782.lbm} & [N/A] & [N/A] & no upstream code \\ \hline
\bottomrule
\end{tabularx}
\par
} 

\end{table}

Our initial plan was to test all 26 benchmarks; three were excluded. 772.marian has been modified so heavily that it bears little resemblance to the original workload, which doesn't even build on AArch64. For 749.fotonik3d and 782.lbm, no public reference workloads or redistributable sources exist, as these codes were contributed directly to SPEC by their authors. We therefore focus on the remaining 23 benchmarks.

\begin{figure}[t]
    \centering
       \includegraphics[width=\columnwidth]{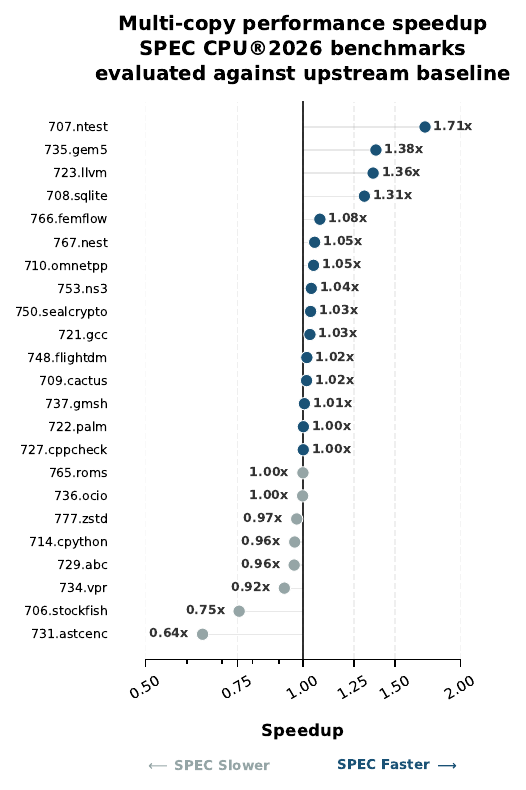}
    \caption{Fully-loaded performance comparison with 192 copies running on AmpereOne. Values greater than 1.0 indicate speedup relative to the original application.}
    \label{fig:192copy_results}
\end{figure}

Execution time was measured for two build configurations within the SPEC harness: the official benchmark binary, and the original sources pre-adaptation. Measurements were collected in single-copy mode (single-threaded performance) and 192-copy mode (throughput under full system load), plotted as S-curves sorted by speedup in \autoref{fig:1copy_results} and \autoref{fig:192copy_results}, and tabulated in \autoref{tab:speedup-data}.

\subsection{Single-Copy Results}

\begin{figure}[t]
    \centering
    \includegraphics[width=\columnwidth]{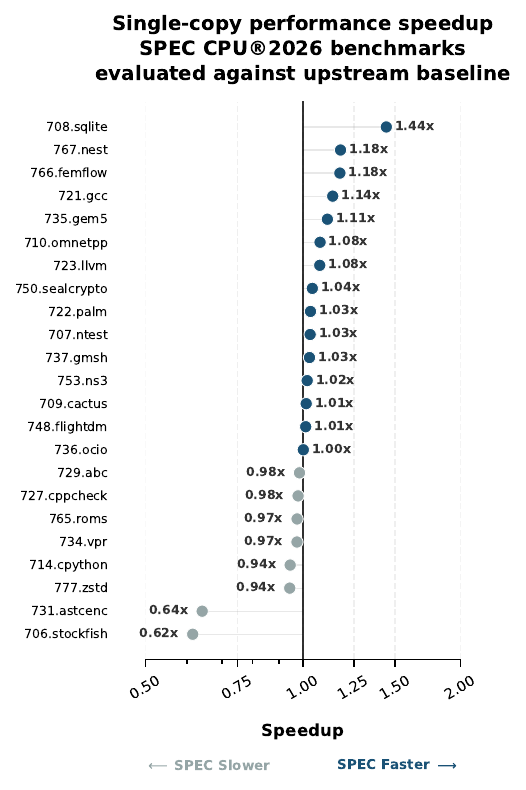}
    \caption{Single-threaded performance comparison with a single copy running on AmpereOne. Values greater than 1.0 indicate speedup relative to the original application.}
    \label{fig:1copy_results}
\end{figure}

\autoref{fig:1copy_results} shows the ratio of SPEC to upstream performance across all 23 benchmarks, falling into three groups. The first group ran notably faster under SPEC (ratios $1.08\times$--$1.44\times$), primarily due to the deliberate removal of I/O and other system-level bottlenecks. The second group clusters around $1.00\times$, within run-to-run variation.  The third group sits well below $1.00\times$, indicating benchmarks where the upstream build outperformed the SPEC binary, which are 706.stockfish, and 731.astcenc; their performance deltas are attributable to the necessary removal of non-portable, hand-tuned assembly or ISA intrinsics. SPEC also suppresses threading/mutex overhead in single-threaded benchmarks, masks non-standard extensions like ``always inline'' in favor of ISO-standard code, and removes randomness that would otherwise contend for the system RNG. In all outlier scenarios, the divergence is a direct consequence of SPEC's methodology, where the benchmark is modified to offer portable code, reproducibility of results, and to keep the spotlight on the computational workload.

\subsection{Multi-Copy Results}

The multi-copy results, shown in \autoref{fig:192copy_results}, provide the clearest evidence of why these adaptations are critical. While most benchmarks maintained similar trends, the results for several stand out. For 707.ntest, 735.gem5, and 723.llvm, the adapted SPEC benchmark becomes far more efficient than the upstream application. This demonstrates that the I/O operations present in the original code create a severe system-level bottleneck under heavy, multi-tasking load, which would have otherwise obscured the CPU's true user-level performance. 706.stockfish improved from $0.62\times$ in single-copy to $0.75\times$ in multi-copy, suggesting that the bottlenecks of scaled execution are consistent in both binaries. As for the benchmarks where the SPEC binary outperformed the upstream build, a possible explanation is resource contention: 766.femflow and 767.nest are positive outliers in single-copy but their delta reduces in multi-copy. Factors such as cache pressure, memory bandwidth, and CPU scheduling overhead may contribute to this effect.

\subsection{Other Factors}
System time in the upstream applications was compared against total execution time; since SPEC benchmarks target user-level compute performance, system time was intentionally reduced. We attribute 8.1 of the 14 percentage-point delta observed in GCC (single-copy) to this reduction; all other upstream applications showed system time well under 1\%. Build tooling can also matter: for Cppcheck, identical \texttt{-O3} flags produced different runtimes depending on toolchain and configuration nuances, suggesting some deltas may stem from unscrutinized Makefile differences.

\section{Benchmark Notes}
\label{notes}
Building each upstream application required benchmark-specific workarounds (e.g., dependency installation, removed command-line flags, or reconciling forked source trees); a full per-benchmark build log is omitted here for space. \autoref{tab:speedup-data} summarizes the root cause of every measured delta; below we detail the largest outliers.

\para{706.stockfish} The original Stockfish 15 \cite{Stockfish_2026} outperformed the SPEC binary by roughly 38\%. Profiling attributes this to \texttt{NNUE::evaluate(...)}, whose execution time dropped by  77\% in the upstream build due to processor-specific intrinsics that SPEC removed for portability.

\para{707.ntest} The ntest build \cite{Petric_ntest_SPEC} originally printed and flushed output after every character; SPEC reduced this to one flush per board, a 100$\times$ reduction in write syscalls, explaining the large multi-copy speedup. Fundamentally this is a discrepancy in how the community uses the ntest tool; typically a user runs one board at a time, not thousands like in the benchmark, so the I/O issues only manifest at scale.

\para{708.sqlite} For SQLite \cite{sqlite_3_33_0}, the dominant runtime delta was isolated to the \texttt{-testset fp} workload and is attributable to SPEC changes in sqlite3.c. Function-level profiling confirms that the difference is concentrated in software floating-point emulation routines, with the largest deltas in \_\_multf3 (43\% of runtime), \_\_divtf3 (7.6\%), and \_\_sfp\_handle\_exceptions (7.0\%). This evidence indicates that increased floating-point conversion/arithmetic overhead drives the performance gap. The benchmark developer confirmed that the \texttt{long double} type was changed to \texttt{double} for portability and fairness, since \texttt{long double} is implementation specific.

\para{723.llvm} For LLVM/Clang \cite{llvm14}, the SPEC benchmark adds a sha512sum to verify the output in memory, to reduce the I/O of writing to disk. A sha512sum on the full text output from the upstream binary matched with the SPEC output, proving equivalence of work. This I/O reduction scales well, leading to a 36\% perf increase at high copy count.

\para{731.astcenc} For ASTCENC \cite{astcencoder2319d9c}, SPEC's removal of assembly-language intrinsics for portability accounts for the observed 35\% slowdown.

\para{735.gem5} In gem5 \cite{gem5_v22_1}, the upstream binary performs an I/O dump of large input configuration files on every run. SPEC suppressed this I/O as it was unnecessary for verification, explaining the upstream binary's slowdown at high copy counts.

\para{721.gcc} We used Ubuntu's cross-compiler package \texttt{gcc-11-x86-64-linux-gnu} rather than the SPEC redistributable tarball, giving what we believe is the most direct comparison to a GCC- build actually used in the field.

\para{766.femflow}
For Femflow \cite{femflow_exadg_2026}, SPEC spliced out unnecessary mutexes and locks since this is a single-threaded benchmark. This explains the 18\% uplift with single-copy.

\para{767.nest}
The NEST simulator\cite{nest_simulator_2021_4739103} uses RNG distributions, which SPEC replaced with `specrand'. Additionally, thread\_local variables were removed. These two explain the perf upside of the benchmark at single-copy.

\section{Conclusion}
\label{sec:conclusion}

This study set out to replace the qualitative critique that ``benchmarks are not real workloads'' \cite{amazon_reinvent} with a quantitative analysis of the fidelity gap between SPEC CPU\textregistered{}2026 and its open-source counterparts. Our findings reveal a nuanced reality: for most benchmarks, the adapted SPEC version correlates closely with the original application, especially single-copy, showing that SPEC's adaptation process often preserves core performance characteristics while achieving portability and determinism.

The most insightful results lie with the outliers. Where SPEC is significantly slower, the cause is typically the removal of non-portable, hand-tuned assembly or ISA intrinsics; where SPEC is significantly faster, the cause is the deliberate reduction of I/O, RNG, and threading bottlenecks. The multi-copy results show these bottlenecks are amplified under parallel load---precisely the conditions under which SPEC's own warning against cross-comparisons is most justified. In both scenarios, the divergence is a direct consequence of the adaptation process, where the benchmark is  modified to isolate the portable algorithms and computational workloads. Therefore, our work does not invalidate SPEC's warning against direct comparisons; rather, it provides the empirical data that \emph{validates} it. The benchmarks can indeed be different, and our findings illustrate precisely why these principled differences are essential for a tool designed to measure computational throughput.

This work provides a data-driven framework for understanding the inherent trade-offs in creating a scientific benchmark: the fidelity gap is a quantifiable consequence of enforcing portability, determinism, and CPU-bound focus, not evidence of a flaw. By measuring it benchmark-by-benchmark, we give the community the context to interpret SPEC CPU 2026 results correctly. appreciating both its close correlation to real-world applications and the specific, principled reasons for divergence. Furthermore, the methodology presented here offers a template for future benchmark development, providing an empirical basis for committees to make data-driven decisions on when a candidate has diverged too far from its real-world counterpart to remain representative.

\bstctlcite{IEEEtran_bst_ctl}

\bibliographystyle{IEEEtranS}
\bibliography{90-citations}

@IEEEtranBSTCTL{IEEEtran_bst_ctl,
  CTLuse_url = {yes}
}

@inproceedings{cpu2026,
      title={{SPEC CPU: The Next Generation}}, 
      author={Mahesh Madhav and others},
      year={2026},
      eprint={2605.01575},
      archivePrefix={arXiv},
      primaryClass={cs.PF},
      publisher = {IEEE},
      booktitle = {Proceedings of the 53rd Annual International Symposium on Computer Architecture},
      series = {ISCA '26},
      location = {Raleigh, NC, USA},
      url = {https://doi.org/10.48550/arXiv.2605.01575}
}

@misc{amazon_reinvent,
      title={{AWS re:Invent 2023 - Compute innovation for any application, anywhere}}, 
      author={Dave Brown},
      year={2023},
      url={www.youtube.com/watch?v=dxm93_qgRzk&t=2210s}, 
}

@manual{AmpereOne_Brief,
  author={Ampere},
  url={https://amperecomputing.com/briefs/ampereone-family-product-brief},
  title={{AmpereOne® Product Brief}},
  year={2024},
  note= {Accessed: May 2, 2026}
}

@misc{warning,
      title={{SPEC CPU 2026 Docs: Are the benchmarks comparable to other programs?}}, 
      author={SPEC},
      year={2026},
      url={https://www.spec.org/cpu2026/Docs/overview.html#Q26}, 
}

@misc{cpuv8_step2,
      title={{Step 2 Rules of the CPUv8 Benchmark Search Program}}, 
      author={SPEC},
      year={2020},
      url={www.spec.org/cpu/cpuv8/#step2} 
}

@misc{reddit,
      title={The fallacy of ‘synthetic benchmarks’}, 
      author={Veedrac},
      year={2020},
      url={www.reddit.com/r/hardware/comments/jvq3do/the_fallacy_of_synthetic_benchmarks} 
}

@misc{statsig,
      title={{Real-world vs benchmark performance: Closing the gap}}, 
      author={{The Statsig Team}},
      year={2025},
      url={www.statsig.com/perspectives/realworld-vs-benchmark-performance} 
}

@inproceedings{how_to_spec,
    author = {v. Kistowski, J\'{o}akim and Arnold, Jeremy A. and Huppler, Karl and Lange, Klaus-Dieter and Henning, John L. and Cao, Paul},
    title = {{How to Build a Benchmark}},
    year = {2015},
    isbn = {9781450332484},
    publisher = {ACM},
    doi = {10.1145/2668930.2688819},
    url = {https://doi.org/10.1145/2668930.2688819},
    booktitle = {Proceedings of the 6th ACM/SPEC International Conference on Performance Engineering},
    pages = {333–336},
    numpages = {4},
    location = {Austin, Texas, USA},
    series = {ICPE '15}
}

@misc{Stockfish_2026,
  author = {{The Stockfish developers}},
  title = {{Stockfish}},
  url = {https://github.com/official-stockfish/Stockfish},
  version = {master},
  year = {2026}
}

@misc{Petric_ntest_SPEC,
  author = {Petric, Vlad},
  title = {{ntest: Othello program}},
  url = {github.com/vladpetric/ntest},
  version = {fc7d6b2},
  year = {2026}, 
  note = {{SPEC version based on commit: fc7d6b26}}
}

@misc{sqlite_3_33_0,
  author = {Hipp, D. Richard},
  title = {{SQLite}},
  version = {3.33.0},
  date = {2020-06-18},
  url = {https://sqlite.org/releaselog/3_33_0.html}
}

@manual{llvm14,
  title        = {{LLVM} Compiler Infrastructure, Version 14.0.0},
  author       = {{LLVM Developer Group}},
  year         = {2022},
  url          = {https://github.com/llvm/llvm-project},
  note         = {Software available from GitHub}
}

@misc{astcencoder2319d9c,
  title        = {{Arm® Adaptive Scalable Texture Compression (ASTC) Encoder}},
  author       = {{Arm Limited}},
  year         = {2021},
  howpublished = {\url{https://github.com/ARM-software/astc-encoder/tree/2319d9c4}},
  note         = {GitHub commit 2319d9c4}
}

@misc{gem5_v22_1,
  author = {{The gem5 Development Team}},
  title = {{The gem5 Simulator, Version 22.1.0.0}},
  year = {2022},
  url = {https://www.gem5.org/project/2022/12/30/gem5-22-1.html}
  }

@misc{nest_simulator_2021_4739103,
  author       = {Hahne, Jan and others},
  title        = {{NEST} 3.7},
  version      = {v3.7},
  url          = {https://github.com/nest/nest-simulator.git},
  note         = {Commit: bf55cc4a}
}

@misc{femflow_exadg_2026,
  author = {Kronbichler, Martin},
  title  = {{FemFlow: A benchmark component of the {exadg} framework}},
  year   = {2026},
  url    = {https://github.com/kronbichler/spec-femflow },
  note   = {Commit: 95acc98}
}

@inproceedings{spec_499,
    author = {Bonebakker, Lodewijk},
    title = {{Comparison of the SPEC CPU Benchmarks with 499 Other Workloads Using Hardware Counters}},
    year = {2008},
    isbn = {9783540698135},
    publisher = {Springer-Verlag},
    address = {Berlin, Heidelberg},
    doi = {10.1007/978-3-540-69814-2_10},
    url = {https://doi.org/10.1007/978-3-540-69814-2_10},
    booktitle = {Proceedings of the SPEC International Workshop on Performance Evaluation: Metrics, Models and Benchmarks},
    pages = {144–153},
    numpages = {10},
    location = {Darmstadt, Germany},
    series = {SIPEW '08}
}

@INPROCEEDINGS{cpu2017_broaden,
  author={Panda, Reena and Song, Shuang and Dean, Joseph and John, Lizy K.},
  booktitle={2018 IEEE International Symposium on High Performance Computer Architecture (HPCA)}, 
  title={{Wait of a Decade: Did SPEC CPU 2017 Broaden the Performance Horizon?}}, 
  year={2018},
  volume={},
  number={},
  pages={271-282},
  url = {https://doi.org/10.1109/HPCA.2018.00032},
  doi={10.1109/HPCA.2018.00032}
}

@misc{cpu2026_comparison,
      title={{SPEC CPU2026: Characterization, Representativeness, and Cross-Suite Comparison}}, 
      author={Ruihao Li and Andrew Jacob and Neeraja J. Yadwadkar and Lizy K. John},
      year={2026},
      eprint={2605.03713},
      archivePrefix={arXiv},
      primaryClass={cs.AR},
      url={https://doi.org/10.48550/arXiv.2605.03713}
}

@inproceedings{serverless,
    author = {Chen, Xinghan and Hung, Ling-Hong and Cordingly, Robert and Lloyd, Wes},
    title = {{X86 vs. ARM64: An Investigation of Factors Influencing Serverless Performance}},
    year = {2023},
    isbn = {9798400704550},
    publisher = {Association for Computing Machinery},
    address = {New York, NY, USA},
    url = {https://doi.org/10.1145/3631295.3631394},
    doi = {10.1145/3631295.3631394},
    booktitle = {Proceedings of the 9th International Workshop on Serverless Computing},
    pages = {7–12},
    numpages = {6},
    location = {Bologna, Italy},
    series = {WoSC '23}
}
~\\
~\\
~\\
~\\
\section*{Acknowledgements}

Given that each workload had its own unique build process, we used OpenAI Codex 5.4 to help navigate and understand the original sources. We felt like we were retracing the steps the SPEC CPU committee took during CPUv8 Step 2 \cite{cpuv8_step2}, dealing with the one-time engineering effort of porting workloads. AI proved particularly valuable understanding the build processes, along with assisting to script the runs and plot the results.

We used Google Gemini 2.5 during the editing process of this paper, for making word choices and refactoring our prose into academic writing. All the text was reviewed and revised by the authors.

We thank the benchmark developers for helping us understand the changes that were needed to adapt applications into benchmarks:  Amin Mohaghegh (vpr), Vlad Petrić (ntest), Alan Mischenko (abc), Prasad Joshi (marian), Fr\'{e}d\'{e}rique Silber-Chaussumier (sqlite), and Ronen Zohar (zstd). Gratitude is also extended to the SPEC CPU committee for scrutinizing this work, providing valuable feedback, and pushing us to root-cause the miscorrelations all the way to completion.

\end{document}